\documentclass[11pt,twoside]{article}
\usepackage{asp2010}
\usepackage{graphicx}
\resetcounters
\begin{document}

\title{Why Galaxies Care about Post-AGB stars}
\author{Hans Van Winckel
\affil{Instituut voor Sterrenkunde, K.\,U.~Leuven, Belgium}
}

\begin{abstract}
Post-AGB stars evolve on a very fast track and hence not many are 
known. Their spectral properties make them, in principle, ideal 
objects to test our theories on the late phases of stellar evolution. 
This has, however, proven much more difficult than anticipated, 
mainly because the morphological, dynamical and chemical diversity 
in Galactic post-AGB stars is very large indeed. Here 
I focus on recent results and touch upon the bright near future 
of post-AGB research.
\end{abstract}

\section{Introduction: Shining Examples}
Very detailed studies of individual objects are quite common in the
astronomical literature of post-AGB stars. Shining examples are, among
others, the Egg Nebula (AFGL 2688), with 747 references in ADS on
29/10/2010; the Calabash Nebula (OH\,231.8 $+$4.2 = QX Pup), 411
references; and the Red Rectangle (HD\,44179), 523 references.  
Despite these thorough studies, there is by no means a consensus on 
how these individual objects are connected by evolutionary channels. 
More generally, the sample of known Galactic post-AGB stars 
\citep{suarez06,szczerba07} is not well connected via detailed 
theoretical evolutionary channels to the asymptotic giant branch 
(AGB) nor to the subsequent planetary nebula (PN) phase
\citep{vanwinckel03}. 

Additionally one of the most popular issues which govern the 
discussion in the research community on the final evolution of low- 
and intermediate-mass stars is the impact of binarity on our global 
understanding of these late stellar evolutionary phases. Confirmed or 
suspected classes of evolved binaries are so prominent (and so poorly 
understood), that our understanding of the final phases of single 
stars may also very well be confused.

During \,their \,evolution \,off \,the \,AGB, \,the \,central \,objects 
\,are \,subject \,to \,drastic changes on a very short timescale 
\citep{vassiliadis93, blocker95}: ~in $\sim$10$^{4}$--10$^{5}$ years, 
the radius changes from several AU down to the final radius of the 
white dwarf (WD). The effective temperature of the central object 
changes from 3000\,K to some 100\,000\,K.  The wind properties of the 
rapidly evolving central stars will also change dramatically, and in 
the last decade it has become clear that the shaping of the 
circumstellar envelope starts very early after the AGB and much earlier 
then was generally acknowledged \citep{balick02,sahai07}. The objects 
will also pass the high-luminosity end of the population II Cepheid 
instability strip, and hence many post-AGB stars of intermediate 
spectral type will pulsate.

Post-AGB stars emit over a very wide range of the electromagnetic 
spectrum and their study requires a multi-wavelength approach. 
The outline of this paper follows the sessions of the conference. 
In Secton~\ref{sect:common}, I highlight their chemical diversity and 
show also that common inhabitants in post-AGB samples may very well be
binaries. In Section~\ref{sect:outthere} I focus on the recently
identified post-AGB stars in extragalactic environments. I end this
contribution by listing some potentially very rewarding future
developments in post-AGB research. In what follows I will use the 
general term ``post-AGB stars" and do not restrict this contribution 
to the objects with resolved circumstellar material, which are \,generally 
\,dubbed \,the \,proto-planetary \,nebulae \,(PPNe).
I will not focus on the shapes and shaping of PPNe and refer the 
interested reader to the proceedings of the recent conference on
Asymmetric Planetary Nebulae \citep{zijlstra11}.

\section{Common Inhabitants}
\label{sect:common}
Post-AGB photospheres bear witness to the total chemical changes
induced by the drege-ups during the whole stellar evolution. A major
change is expected to occur during the AGB phase by the third dredge-up 
phenomenon. During the relaxation period after a thermal pulse, 
products of the internal nucleosynthesis can be brought to the surface 
of the star, while at the same time fueling the synthesis of heavy 
elements by inducing protons into the intershell. The formation of 
$^{13}$C and the $^{13}$C($\alpha$,n)$^{16}$O reaction can initiate 
the slow neutron-capture reaction chains (the $s$-process) deep in
the stellar interior. For more massive stars and/or at the hotter
parts of the He shell-flash, the $^{22}$Ne neutron source will become 
active (see Karakas in these proceedings). 
Synthesis by the $s$-process is thought to be an important contributor 
to the cosmic abundances past the iron peak, and solar-type stars are 
very important contributors to this, as well as to the total carbon 
and nitrogen enrichment of the parent galaxy. 

This rich nucleosynthesis is, however, only detected in a small subset 
of Galactic post-AGB stars, namely the post-Carbon stars 
\citep[e.g.][]{vanwinckel00, reyniers02, reddy02,reyniers02thesis}. 
These also show a characteristic dust feature in their infrared spectra 
at 21 $\mu$m, which is only detected in circumstellar dust envelopes of
carbon-rich post-AGB stars \citep[e.g.][]{kwok89,hrivnak10}. The
carrier of this feature is still a matter of debate.

\articlefiguretwo{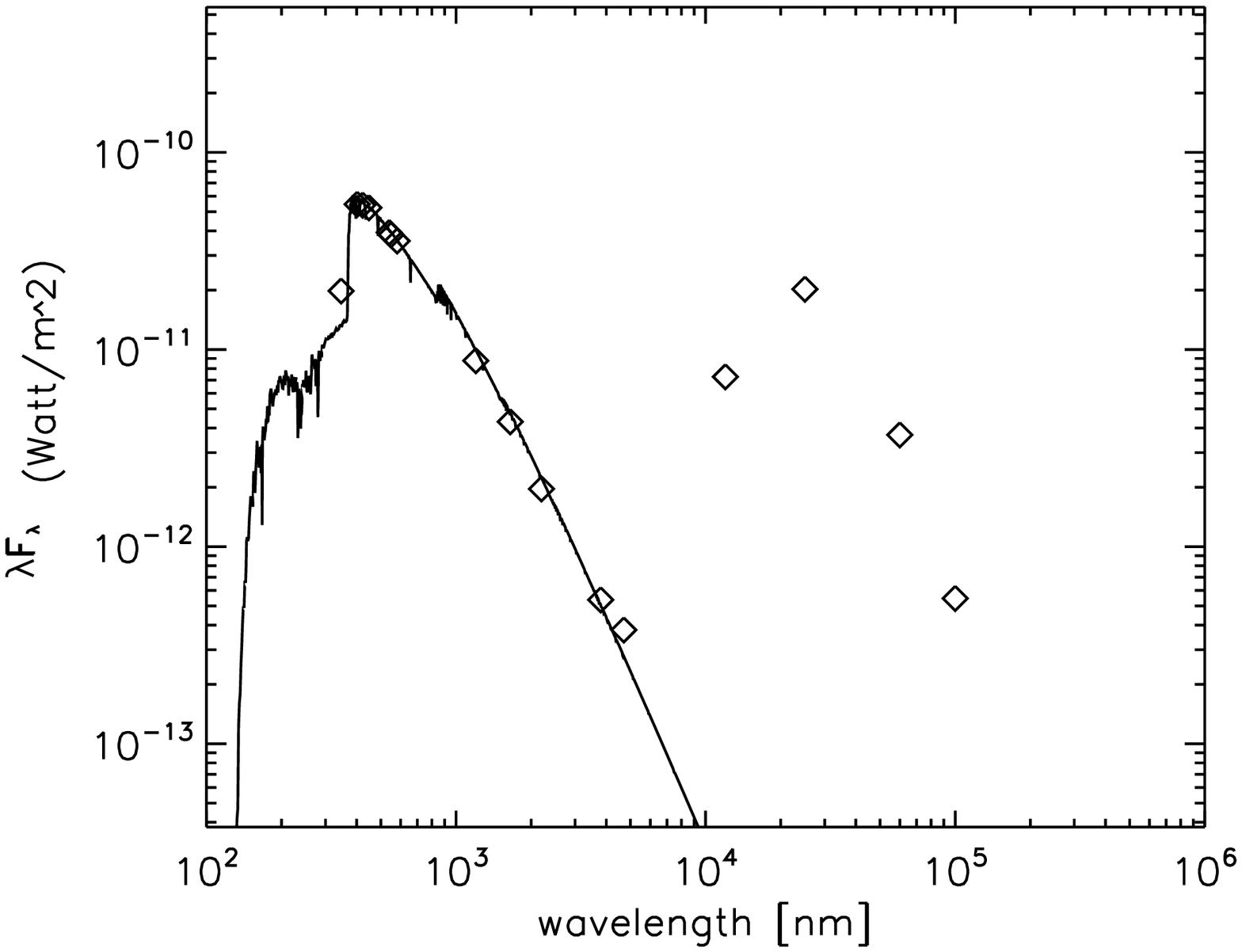}{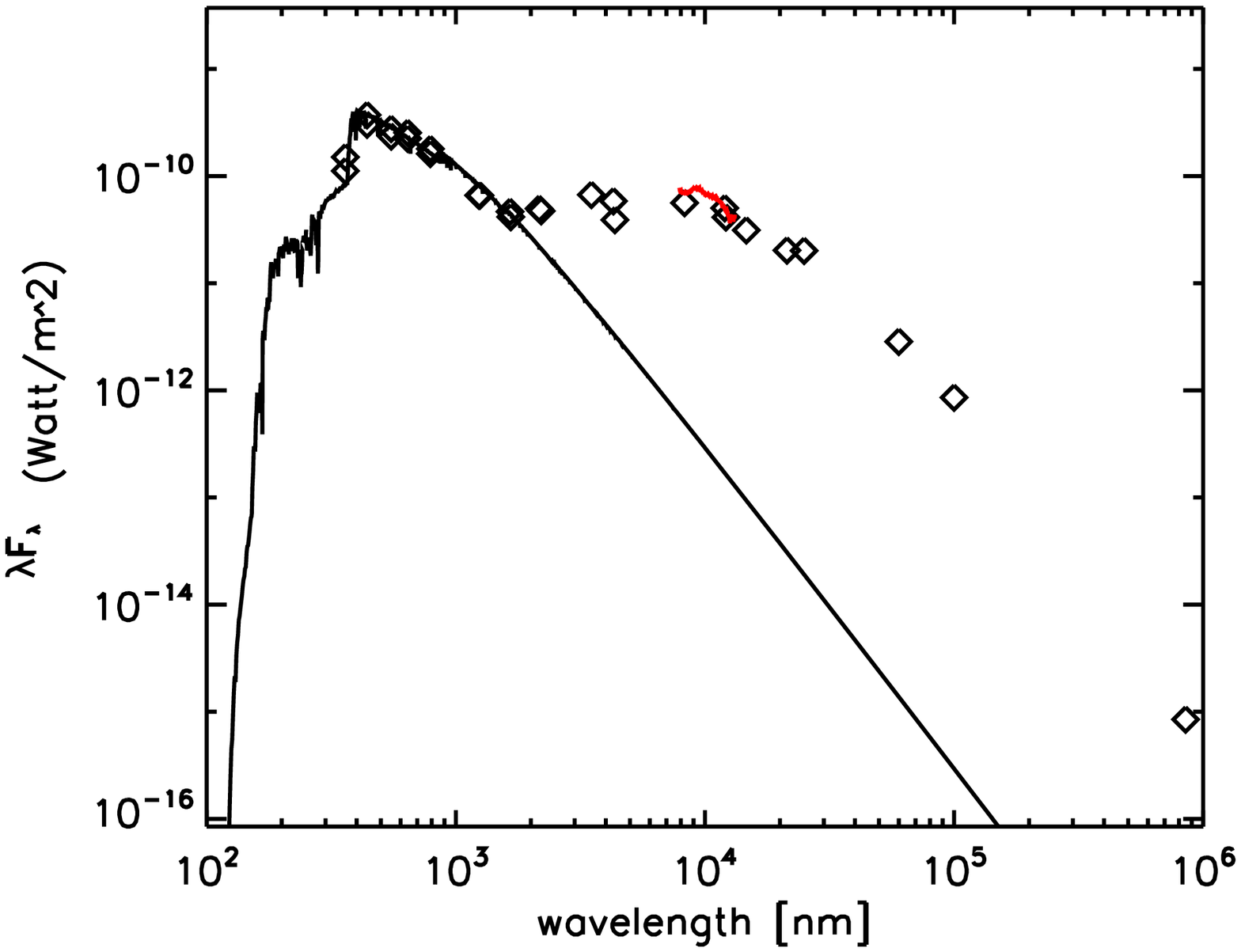}{galactic}{ 
  Spectral energy distributions of two clearly different post-AGB stars. 
  The full lines are the appropriate Kurucz atmospheric models. 
  The points indicate dereddened broad-band fluxes. {\it Left}: 
  HD~187885, a carbon and $s$-process enhanced post-AGB star 
  \citep*{vanwinckel96b}.  {\it Right}: ~the SED of IRAS 08544--4431
  \citep{maas03}, a binary with a resolved compact circumstellar disc
  \citep{deroo07}. The flux at sub-mm wavelengths indicate the presence
  of large circumstellar grains \citep{deruyter05}.}

Some post-AGB stars are the most $s$-process enriched objects known to 
date \citep[e.g.][]{reyniers04}, while others are not enriched at all. 
This dichotomy is very strict, in the sense that mildly enhanced objects 
do not exist in current Galactic post-AGB samples (except for a few
rather atypical objects). This is not expected, as a more gradual
transition between non-enriched and enriched photospheres will occur, 
if the transition from an O-rich AGB to a C-rich AGB star happens
over many thermal pulses. 
Stellar evolutionary AGB models cannot predict very well the minimum 
initial stellar mass for the third dredge-up to occur 
(see Stancliffe in these proceedings), and the dependence of the
third dredge-up phenomenon on the metallicity, initial mass, and 
luminosity still remains very uncertain.  
Examples exist of very similar post-AGB stars (in metallicity,
spectral type, infrared excess, etc.) with totally different
photospheric abundance patterns. These results clearly illustrate that
the third dredge-up phenomenon, and the associated partial mixing of
protons invoked to explain the $s$-process nucleosynthesis, is still 
far from being fully understood.

Binary objects tend to have a totally different photospheric
composition than suspected single objects, showing some degree of
depletion of refractory elements in their photosphere: elements with
a high dust condensation temperature are systematically under-abundant
\citep[e.g.][]{giridhar05, maas05,reyniers07b}.  In
almost all depleted post-AGB objects, there is observational evidence
that a stable circumbinary disc is present \citep{deruyter06}  and we
can use the typical infrared signatures of these circumbinary discs 
(see Figure~\ref{galactic}) to discriminate between probable single 
stars and evolved binaries.  The disc seems to be a prerequisite to 
obtain the photospheric depletion patterns by accretion of gas, 
cleaned from refractories by dust formation \citep*{waters92}.
Interferometric studies confirm the very compact nature of the
circumstellar material \citep{deroo06, deroo07} and infrared
spectroscopy shows the very high degree of processing of the dust 
grains in the discs \citep{deruyter05,gielen07,gielen08} 
(see also Gielen, this conference).

Our radial velocity program is still on-going, but we indeed can
confirm the suspected high binary rate: for non-pulsating objects with
energy distributions pointing to a disc, a binary rate of 100\% was 
found \citep{vanwinckel09}. The companion stars are likely unevolved
main-sequence stars, which do not contribute significantly to the
energy budget of the objects. The orbits are now not in contact, but
they are too small to have accommodated an AGB star.  Our radial
velocity monitoring program is still continuing with our new 
{\small HERMES} high-resolution spectrograph \citep{raskin10}. 
After a few more years of monitoring, this program will allow a good 
statistical overview of the orbital properties of these evolved 
binaries.

The global picture that emerges is that a binary star evolved in a
system which is too small to accommodate a full grown AGB star. During
a badly understood phase of strong interaction, a circumbinary dusty
disc was formed, but the binary system did not suffer dramatic spiral-in. 
What we observe now is an F--G post-AGB supergiant in a binary
system, which is surrounded by a circumbinary dusty disc.  The objects
were likely truncated during their ascent on the AGB
branch. Observational hints for this truncated AGB evolution is that
in all objects, the circumstellar dust in the disc is oxygen-rich (see
Gielen, this conference). The formation of the disc occurred when the
object was still an M star on the AGB. Subsequent thermal pulses may
have occurred. In some objects double chemistry is detected in which 
the C-rich component is mainly limited to PAH emission 
\citep[e.g.][]{gielen09a}. This does not necessarily mean that the 
central object became a carbon star, as PAH emission is also found in 
environments where dissociation of CO can liberate C atoms. Objects 
like the Red Rectangle \citep[e.g.][]{witt09} and HR\,4049 
\citep{geballe89} display a richer C-rich dust and gas component.

In the Galaxy, post-AGB stars with observational evidence for a disc
are a very significant fraction of all known post-AGB stars
\citep{vanwinckel03} and they are certainly common inhabitants.

\section{Out There}
\label{sect:outthere}

So far relatively little focussed work can be found on extragalactic
post-AGB stars \citep[e.g.][]{wood01,kraemer06, reyniers07a}. The 
first systematic searches were based on the data from microlensing
projects. The light curve databases were scanned to find variability 
similar to luminous pulsating population II Cepheids such as RV\,Tauri 
stars \citep{alcock98}.

\articlefigurefour{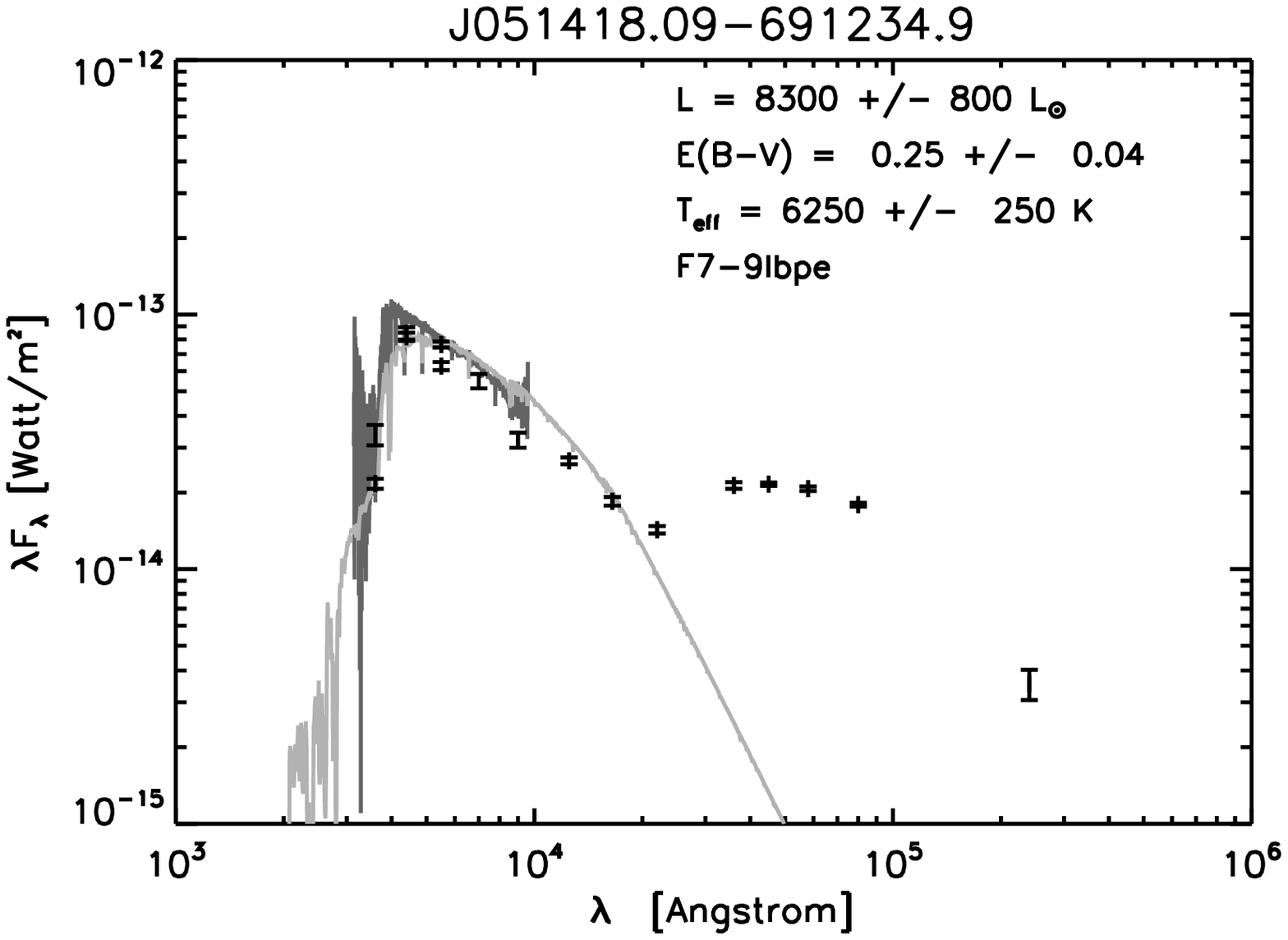}{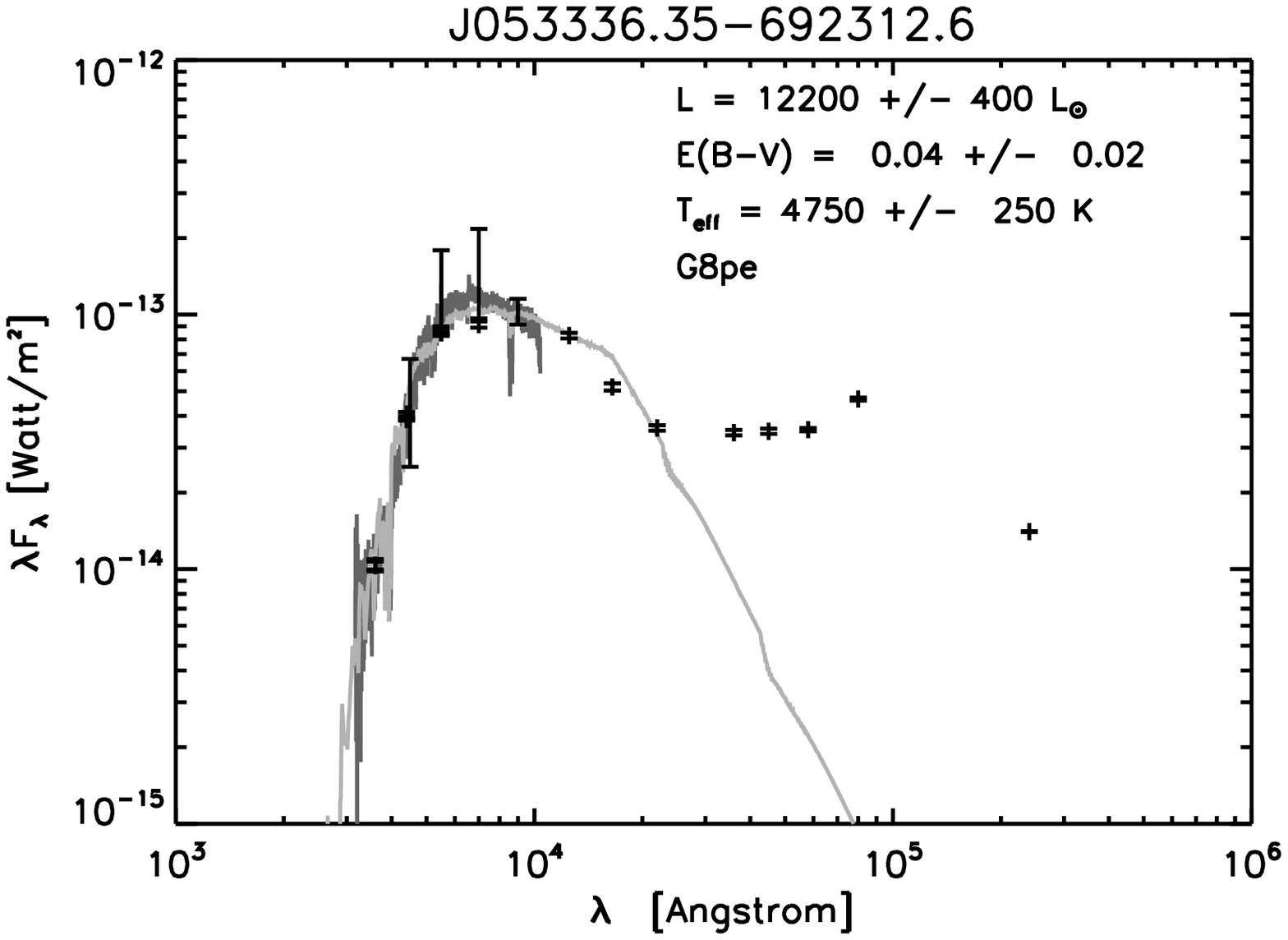}
{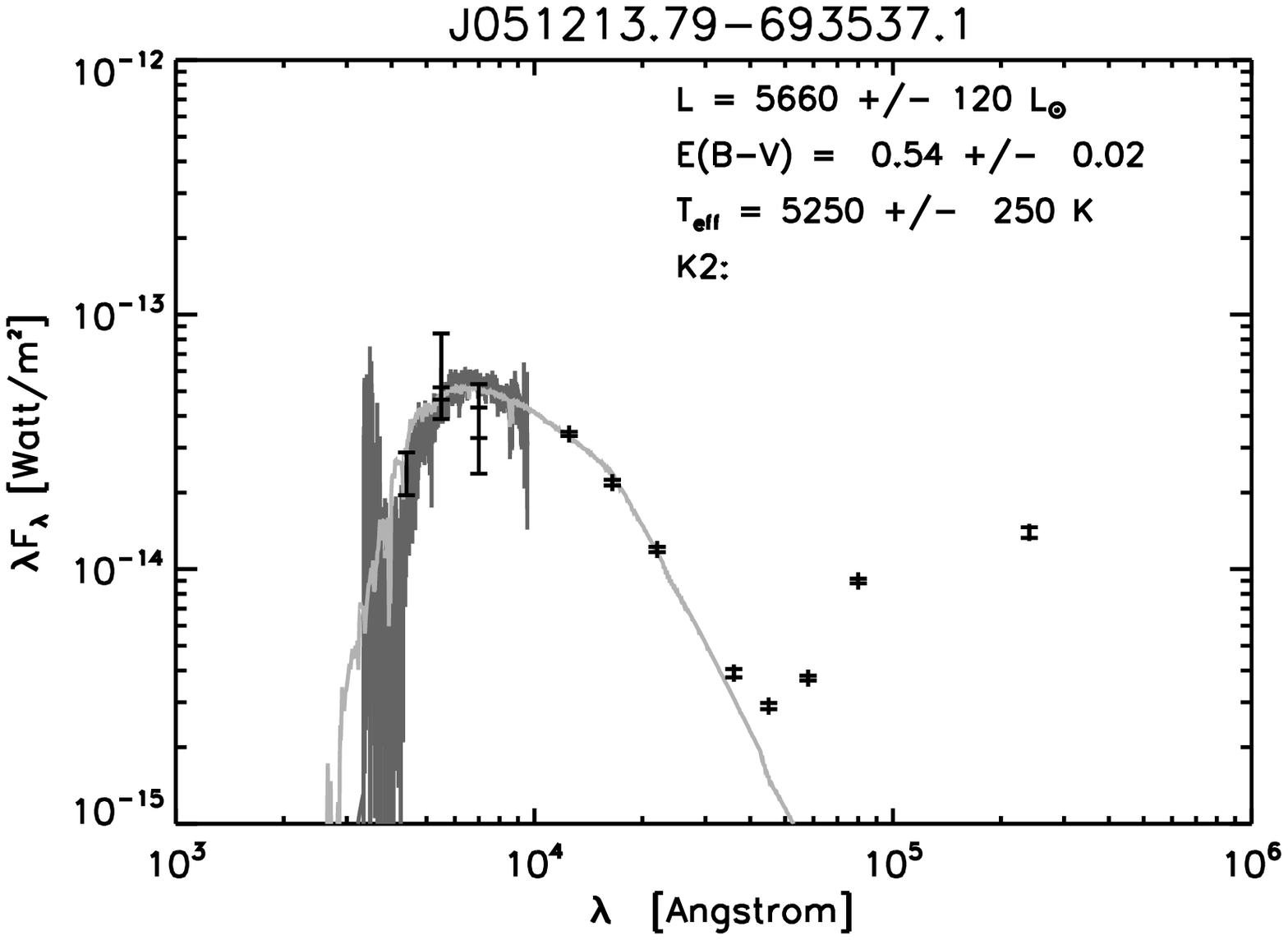}{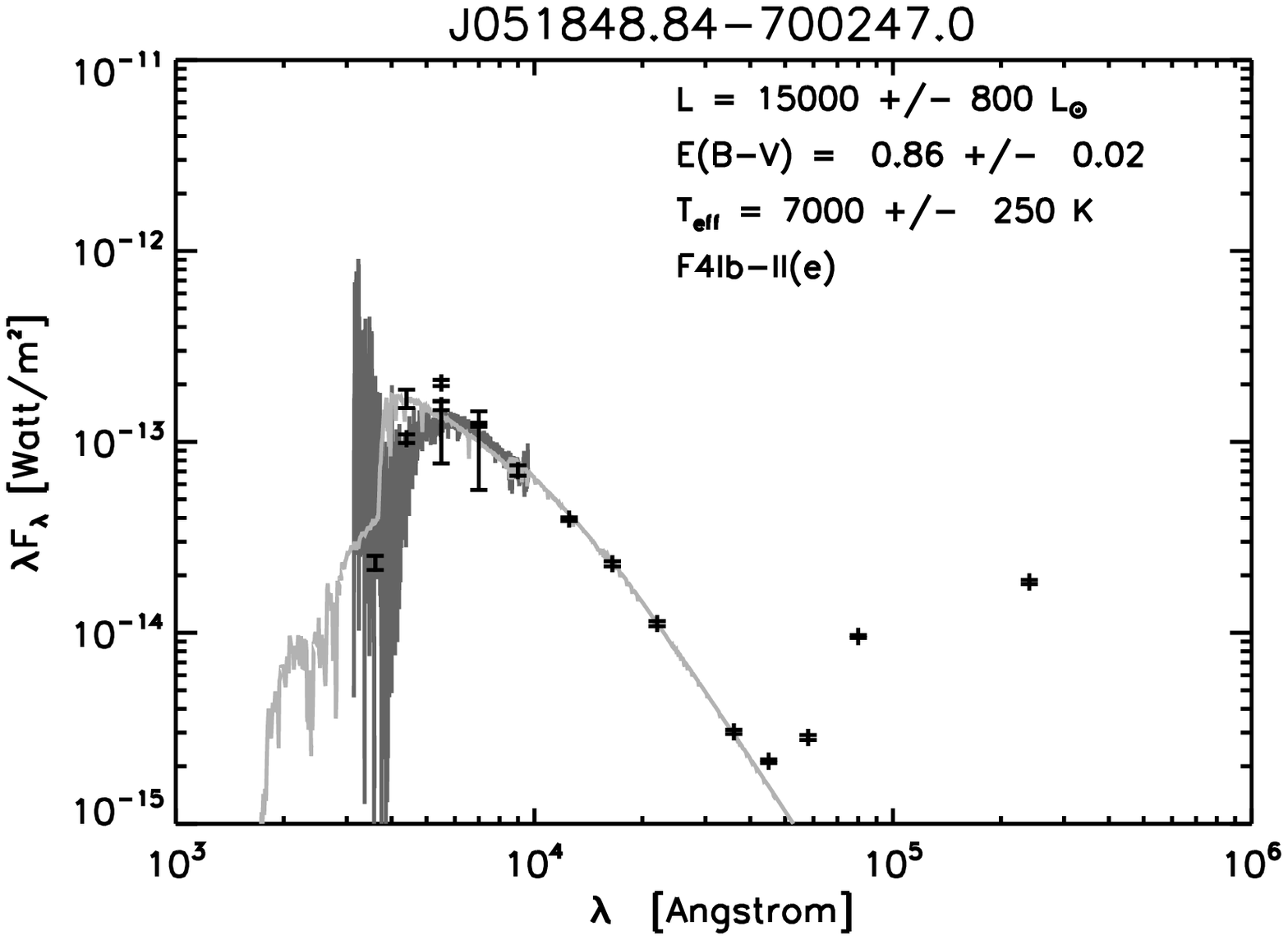}{sedlmc}{SEDs of post-AGB 
objects in the LMC (van Aarle 2010, submitted). The top panels 
show SEDs of disc sources while the bottom ones are more compliant 
with a detached shell.}

A major step forward is offered by the large IR surveys which are deep
and accurate enough to resolve individual extragalactic objects.  The 
data release of the infrared SAGE--{\sl Spitzer} survey of the LMC 
includes about 6.4 million sources detected with IRAC (several 
mid-infrared bands up to 8 $\mu$m) and 60\,000 objects detected at 
24 $\mu$m \citep{meixner06}.

The systematic sample selection of post-AGB candidate stars in the 
LMC by van Aarle (submitted) is based on the second SAGE data 
release (September 2009), and we searched for luminous, optically 
bright stars with infrared colours indicative of a past history of 
dusty mass loss.

To \,discriminate \,between \,genuine \,post-AGB \,stars \,and \,other 
\,objects \,with
\,IR excesses such as luminous young stellar objects, PNe, dusty 
supergiants, etc., the integrated luminosity is used as a selection 
criterion as well. Moreover, to probe the spectral type of the central 
objects, a low-resolution optical spectral survey was performed 
at Siding Spring, Australia, and at SAAO (South Africa).  
The final catalogue consists of 1780 good candidate post-AGB stars 
of which 66 are well characterised with low-resolution spectra at 
this stage. About half of this sample show indications of a 
circumstellar disc rather than an expanding outflow! To prove that 
these are binaries will indeed be an observational challenge.

For the SMC, with its lower global metallicity, the third dredge-up 
enrichment is predicted to be stronger, as witnessed by the
low-luminosity tail of the intrinsic carbon star luminosity function
of the SMC \citep[e.g.][]{lagadec07}. The sample selection of the SMC 
was based on the {\sl Spitzer} S3MC survey \citep{bolatto07}
in a very similar way as was performed on the LMC data. So far, 
low-resolution optical data for 34 of the post-AGB candidates in the 
SMC have been registered. 

Our observing proposal (PI: Peter Wood) for a full low-resolution 
spectral survey of all good candidates, using the wide-field, 
multi-object spectrograph at Siding Spring, has been accepted. The
low resolution spectra will be used to constrain the spectral type 
of the central star so that SEDs can be determined with good accuracy. 

\section{Perspectives}
\label{sect:perspective}

Post-AGB research has often concentrated on very detailed
astrophysical investigations of bright and often spectacular Galactic
stars.  These detailed studies of nearby objects are now expanded into
the far-infrared with the {\sl Herschel} satellite  (see the papers
of Groenewegen, Bujarrabal, and Wesson in these proceedings). 
The harvest of this unique facility is, at the time of the writing, 
still in its infancy.  Theoretical interpretation in terms of stellar
evolution has always been hampered, however, by poorly
constrained distances. It is not often clear whether the very
thoroughly studied objects are representative of a large population
of stars or mere odd-balls which populate a very specific region in
the complete parameter space.

With the advent of large-scale infrared surveys of nearby
galaxies, the exploitation of unique extragalactic samples of 
post-AGB stars has yet to begin. The complete catalogues 
offer strong potential in the study of mass-dependent
properties of post-AGB stars. 

The potentials are illustrated in Figures~\ref{j050631_spec} and
\ref{j050631}, where sample spectra are shown together with the
preliminary results of our photospheric chemical study.  The strong
spectral impact of $s$-process enhancement on the photospheric spectra
of these objects is clear. Unlike AGB stars (see Abia, these
proceedings), these spectra are dominated by atomic lines and 
overall are much cleaner than AGB photospheric spectra.  Detailed 
and accurate photospheric abundances can therefore be traced, from 
CNO to the heavi\-est elements well beyond the Ba peak. A detailed 
comparison between theoretical enrichment models and the chemical 
composition of a large sample of post-AGB stars with well-constrained 
luminosities will be very rewarding. It will allow us to put stringent 
constraints on the dredge-up occurrence in function of metallicity and 
luminosity, as well as on the $s$-process nucleosynthesis itself and 
its associated mixing uncertainties (see Karakas, these proceedings).
The study of the connection between the large number of post-AGB stars 
and the PNe of the LMC and SMC (see Stanghellini, these proceedings)
will become possible in the near future when the former catalogue is
evaluated more completely with the low-resolution spectral survey. \\

\articlefigure{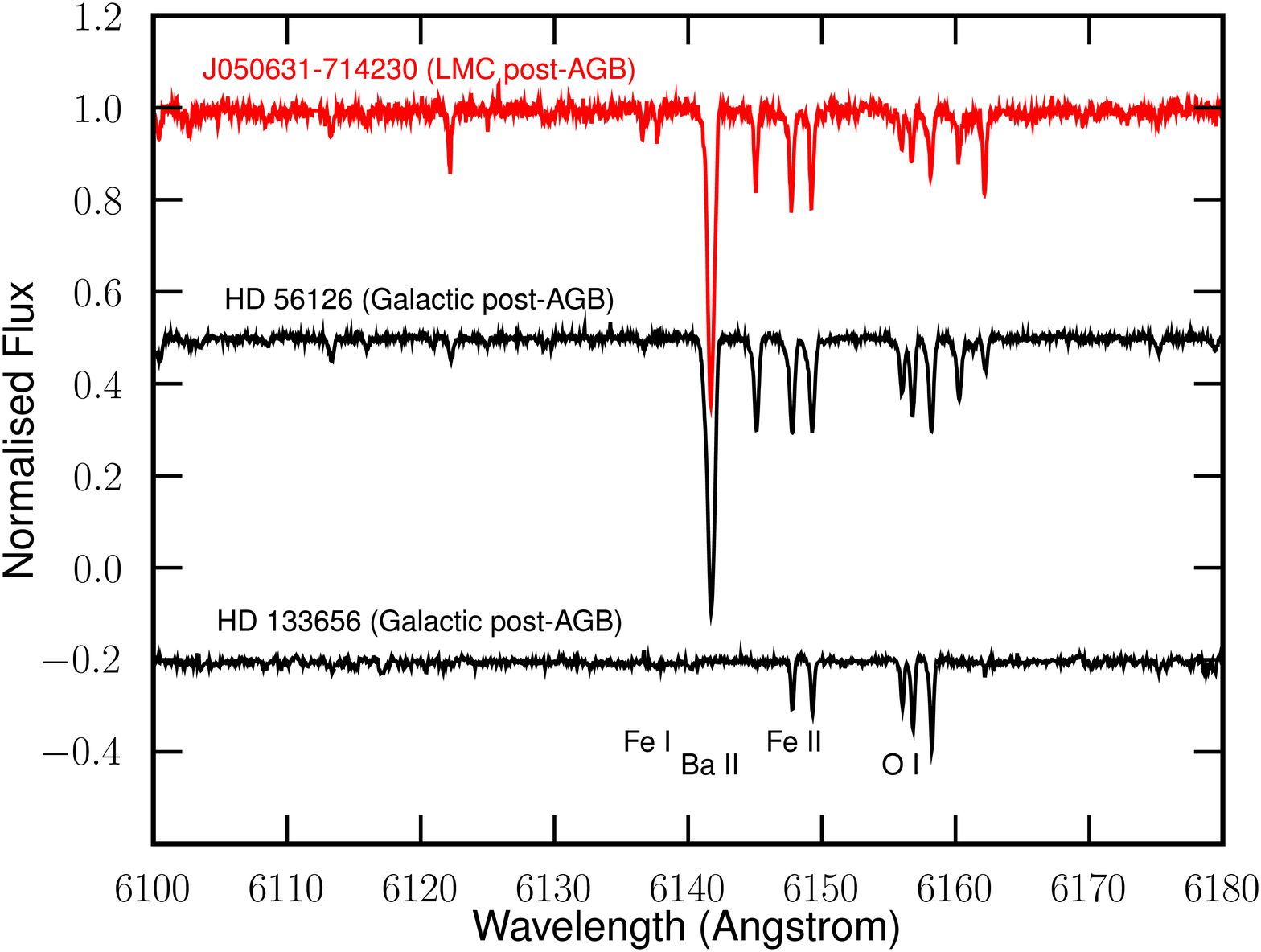}{j050631_spec}{
  UVES spectra of three post-AGB stars in the region of the 6141 \AA\ 
  line of Ba\,{\sc ii}.  The {\it top} spectrum is the LMC star 
  J050631--714230, the SED of which is shown in Fig.~\ref{j050631}. 
  The {\it middle} spectrum shows HD\,56126 \citep{hony03}, a 
  Galactic carbon and $s$-process enriched post-AGB star. 
  The {\it bottom} panel shows a non-enriched post-AGB stars 
  HD~133656 \citep*{vanwinckel96a} of similar spectral type and 
  metallicity.}

\articlefiguretwo{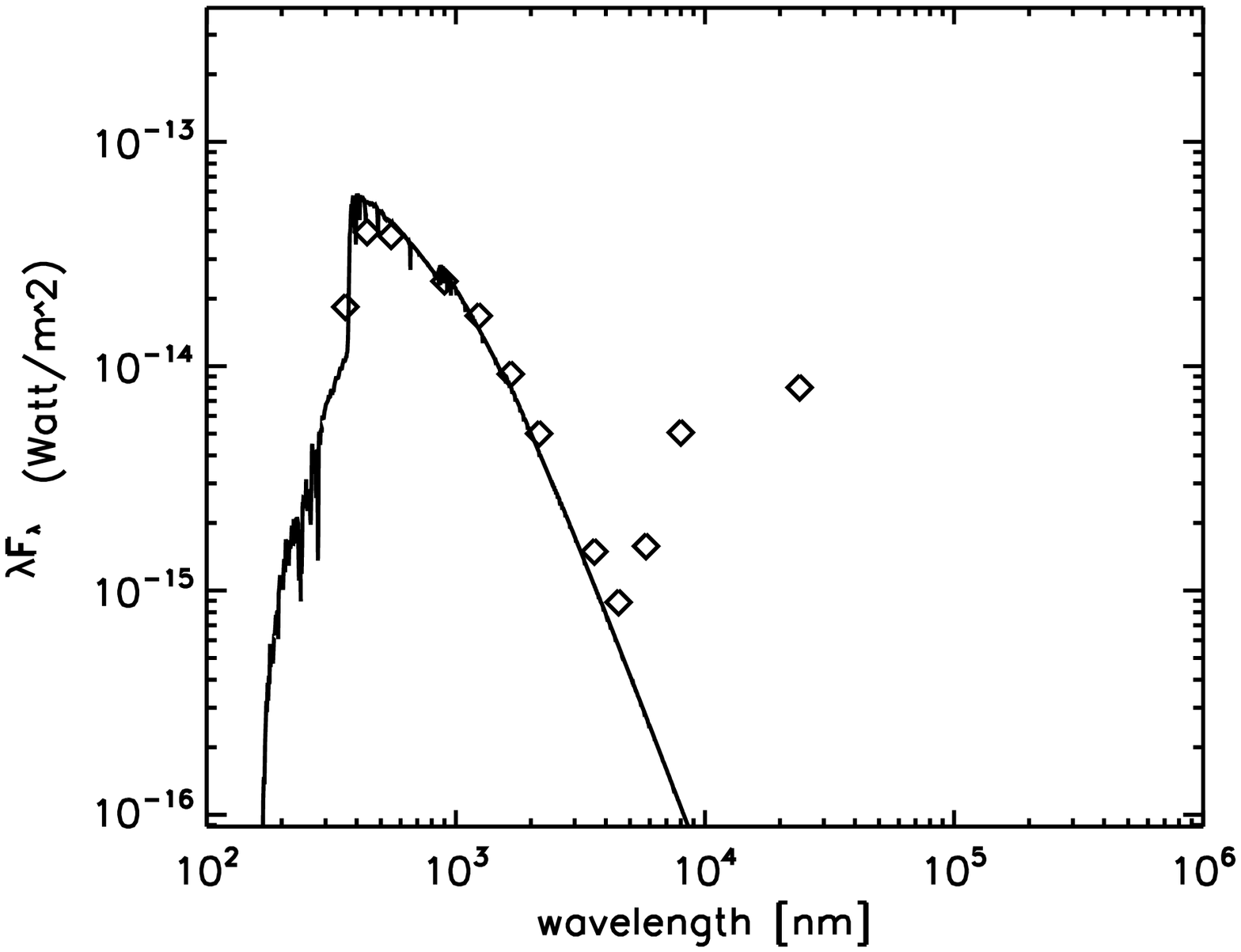}{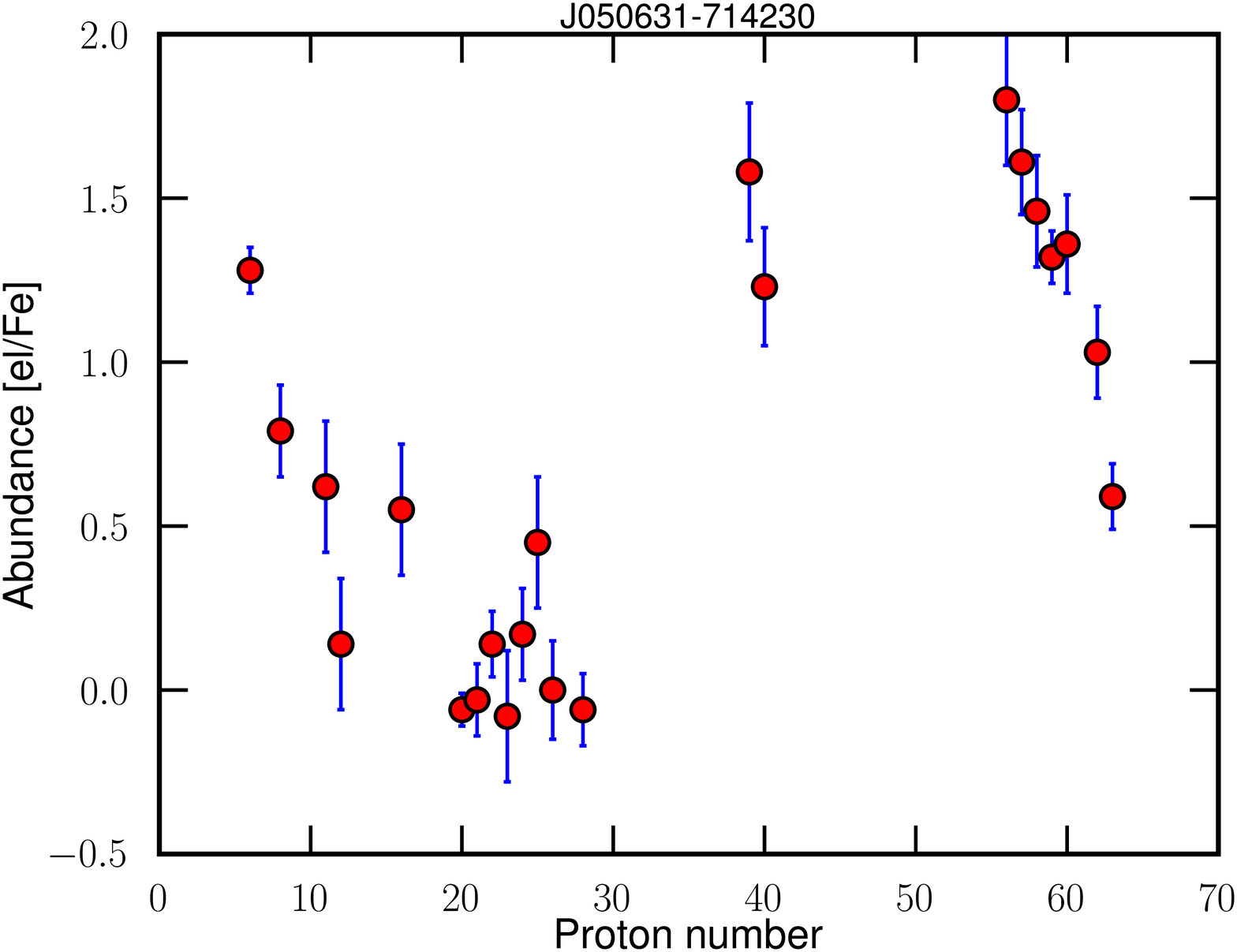}{j050631}{
  {\it Left}: the SED of J050631--714230, a post-AGB star in the LMC 
  with a detached shell (van Aarle 2010, submitted). {\it Right}: the 
  photospheric chemical composition of this strongly $s$-process 
  enriched object of low initial metallicity ([Fe/H] = --1.4) 
  (van Aarle 2011, in prep.).}

Additionally, the spectral complement to the SAGE survey,
SAGE--{\small SPEC} \citep{kemper10}, and the spectral survey of the 
point-sources \citep{woods10} will allow detailed comparison between 
the circumstellar gas and dust features with the often bright central 
star.

It is now generally acknowledged that post-AGB sources with discs are
associated with binarity of the central stars. The discs have a longer
infrared lifetime and the observed feedback from the disc onto the
central star may very well prolong its post-AGB lifetime. 
The orbits obtained so far in the Galactic sample require that the 
central objects were subject to severe binary interaction processes 
when the primary was at giant dimensions. 
The discs in post-AGB stars are always associated with oxygen-rich 
dust. It did come as a surprise that stratified and very compact discs 
have been found around PNe as well \citep{chesneau07,chesneau10} 
but it is as yet very unclear whether these discs are relics of the 
discs detected around post-AGB binaries or are much more recently formed.  
Disc creation and disc evolution should therefore be an important 
ingredient in any binary evolution model with an evolved low- to 
intermediate-mass object. 

The periods and eccentricities found in post-AGB binaries are not too 
different from that found in Ba stars \citep{jorissen99}. These 
extrinsically enhanced systems harbour a WD which polluted its now seen 
companion.  Despite the orbital similarity, it is very unlikely that 
there is an evolutionary connection between the post-AGB stars and the 
Ba star family as no $s$-process enrichment nor carbon-rich circumstellar 
dust has generally been found in post-AGB stars. The study of the
evolutionary connection between post-AGB binaries and the zoo of
other binaries with evolved components --- Ba stars, symbiotic stars,
CH stars \citep[e.g.][]{jorissen99}, sdBs, bipolar PNe, CVs, 
sequence~E AGB stars \citep*[e.g.][and these proceedings]{nicholls10}, 
spiralled-in PNe, etc. --- is an additional challenge for many years 
to come.

\acknowledgements The LMC-SMC post-AGB project includes collaboration
with Els van Aarle, Tom Lloyd Evans, Peter Wood, Toshiya Ueta, Devika
Kamath, and Clio Gielen.

\bibliography{b_vanwinckel}

\end{document}